\documentclass[aps,prb,twocolumn,superscriptaddress,showpacs,floatfix,reprint]{revtex4-1}

\usepackage{amsmath}
\usepackage{amssymb}
\usepackage{graphicx}

\newcommand{\cE}{{\cal E}}
\newcommand{\Vdc}{{V_{\rm dc}}}
\newcommand{\Vac}{{V_{\rm ac}}}
\newcommand{\cS}{{\cal S}}
\newcommand{\mb}[1]{{\bf #1}}
\newcommand{\spann}{{\rm span}}

\newcommand{\ket}[1]{{\vert #1 \rangle}}
\newcommand{\bra}[1]{{\langle #1 \vert}}
\newcommand{\scp}[2]{{\langle #1 \vert #2 \rangle}}

\newcommand{\ga}{\alpha}

\newcommand{\Tr}{{\rm Tr}}

\begin{document}

\title{Electron and electron-hole quasiparticle states in a driven quantum contact}

\author{Mihajlo Vanevi\' c}%
\affiliation{Department of Physics, University of Belgrade,
Studentski trg 12, 11158 Belgrade, Serbia}

\author{Julien Gabelli}
\affiliation{Laboratoire de Physique des Solides, Univ. Paris-Sud,
CNRS, UMR 8502, F-91405 Orsay Cedex, France}

\author{Wolfgang Belzig}
\affiliation{Fachbereich Physik, Universit\" at Konstanz, D-78457
Konstanz, Germany}

\author{Bertrand Reulet}
\affiliation{D\'epartement de physique, Universit\'e de Sherbrooke,
Sherbrooke, Qu\'ebec, J1K 2R1, Canada}

\date{\today}

\begin{abstract}
We study the many-body electronic state created by a time-dependent
drive of a mesoscopic contact. The many-body state is expressed
manifestly in terms of single-electron and electron-hole
quasiparticle excitations with the amplitudes and probabilities of
creation which depend on the details of the applied voltage. We
experimentally probe the time dependence of the constituent
electronic states by using an analog of the optical Hong-Ou-Mandel
correlation experiment where electrons emitted from the terminals
with a relative time delay collide at the contact. The electron wave
packet overlap is directly related to the current noise power in the
contact. We have confirmed the time dependence of the electronic
states predicted theoretically by measurements of the current noise
power in a tunnel junction under harmonic excitation.
\end{abstract}




\maketitle


Recent years have seen a tremendous experimental and theoretical
progress in the emerging field of electron quantum
optics.\cite{bocquillon_electron_2014} Following the example of
optics, the quantum nature of electronic transport has been
demonstrated in electronic Mach-Zehnder
interferometer\cite{ji_electronic_2003} and Hanbury
Brown-Twiss\cite{henny_fermionic_1999,bocquillon_electron_2012,ubbelohde_partitioning_2015}
and
Hong-Ou-Mandel\cite{liu_quantum_1998,neder_interference_2007,bocquillon_coherence_2013}
intensity correlation experiments. Although quantum optics with
electrons is in general analogous to the one with photons, there are
important distinctions between the two due to differences in particle
statistics, vacuum state (Fermi sea vs. photonic vacuum), interaction
between electrons, decoherence, etc. In particular, a simple constant
voltage source can act as a single-electron
turnstile\cite{thibault_pauli-heisenberg_2015} due to the Fermi
statistics which is responsible for regular emission of electrons on
a time scale $h/eV$, where $e$ is the electron charge, $h$ is the
Planck constant, and $V$ is the dc voltage drop over the conductor.

A step forward towards electron quantum optics has been made recently
with the realization of {\em on-demand} electron sources%
\cite{bocquillon_electron_2012,ubbelohde_partitioning_2015,bocquillon_coherence_2013,%
gabelli_violation_2006,feve_on-demand_2007,dubois_minimal-excitation_2013,%
jullien_quantum_2014,gabelli_shaping_2013}
which can create single- to few-particle excitations.%
\cite{keeling_minimal_2006,*ivanov_coherent_1997,*levitov_electron_1996,%
keeling_coherent_2008,moskalets_quantized_2008,olkhovskaya_shot_2008,%
parmentier_current_2012,haack_glauber_2013,moskalets_first-order_2015}
This facilitates the full control of the quantum state of electrons
in mesoscopic conductors and the dynamical control of elementary
excitations using suitably tailored voltage
pulses.\cite{vanevic_control_2012}
In particular, time-dependent drive creates quasiparticle excitations
in the Fermi sea that are single-electron and electron-hole pairs
whose number and probability of creation depend on the shape and the
amplitude of the applied
voltage.\cite{vanevic_elementary_2007,*vanevic_elementary_2008}
Lorentzian pulses $V(t)$ of a quantized area $\int eV(t)dt/h=N$ ($N$
is an integer) are special as they create $N$ electrons above the
Fermi level leaving the rest of the Fermi sea
unperturbed.\cite{keeling_minimal_2006,*ivanov_coherent_1997,*levitov_electron_1996}
Experimentally, the presence of electron-hole pairs in the system can
be seen in the zero-frequency photon-assisted current noise power
which is increased with respect to the dc noise level.%
\cite{SchoelkopfKozhevnikovProberRooksPRL80-98,KozhevnikovProberPRL84-99,%
ReydelletRocheGlattliEtienneJinPRL90-03} More recently, quantum noise
oscillations have been observed in a driven tunnel
junction\cite{gasse_observation_2013} and noise spectroscopy using a
more complex biharmonic voltage drive has been carried out
approximating Lorentzian pulses.\cite{gabelli_shaping_2013} The
creation of single-electron excitations has been realized
experimentaly\cite{dubois_minimal-excitation_2013} and the resulting
quantum states have been reconstructed using the quantum state
tomography.\cite{jullien_quantum_2014}

Even though the progress has been exceptional,
a fundamental question remains: What is the many-body electronic
state created in the Fermi sea by a 
voltage drive? In this article we find that the many-body state is:
\begin{equation}\label{eq:Psi}
\ket{\Psi}=\hat C^\dagger \prod_k \left(
    \sqrt{1-p_k} + i\sqrt{p_k}\hat A_k^\dagger \hat B_k
\right)\ket{F}.
\end{equation}
Here, $\hat C^\dagger = \int d\cE {\rm v}^*(\cE)\hat c^\dagger(\cE)$
is the creation operator of a single-electron quasiparticle state,
$\hat A_k^\dagger = \int d\cE {\rm u}_+^{k*}(\cE) \hat
c^\dagger(\cE)$ and $\hat B_k = \int d\cE {\rm u}_-^k(\cE) \hat
c(\cE)$ are the operators that create electron and hole from the
electron-hole pair, $\hat c^\dagger$ ($\hat c$) are the electron
creation (annihilation) operators, and $\ket{F}$ is the filled Fermi
sea. [We have assumed, for simplicity, that there is one
single-electron quasiparticle created per period, which is the case
for $e\Vdc/\hbar \omega = 1$ where $\Vdc$ is a dc voltage component
and $\omega$ is the frequency of the drive.] In addition to a
single-electron excitation, there is a number of electron-hole pairs
created in the system (labelled by $k=1,2,\ldots$) due to the ac
voltage component. The probabilities of the electron-hole pair
creations $p_k$ and the single-electron and electron-hole
quasiparticle amplitudes ${\rm v}(\cE)$ and ${\rm u}_\pm^k(\cE)$
depend on the properties of the applied voltage. For optimal
Lorentzian drive, there are no electron-hole pairs created ($p_k=0$)
and the state has only single-electron excitations, as expected.

Quasiparticle amplitudes $\rm v$ and ${\rm u}_\pm^k$ in
Eq.~\eqref{eq:Psi} also give the time-dependent probabilities of
single-electron and electron-hole pair creations, $|\rm v(t)|^2$ and
$|{\rm u}_\pm^k(t)|^2$. The time dependence of the wave functions
that constitute the many-body state in Eq.~\eqref{eq:Psi} can be
probed by an electronic analog of the optical Hong-Ou-Mandel
correlation experiment,\cite{hong_measurement_1987,*ou_further_1989}
where electrons emitted from two terminals with a relative time
delay collide at the contact.%
\cite{dubois_integer_2013,grenier_single-electron_2011} When the wave
packets arrive at the contact simultaneously, their transmission in
the same output channel is blocked by the Pauli principle, which
suppresses the current fluctuations. The magnitude of the noise
suppression is proportional to the wave-packet overlap at the
contact. In the present paper, we have performed this experiment and
measured current noise power in a tunnel junction driven by harmonic
time-dependent voltage. We have found that the correlation noise as a
function of a time delay is in agreement with the theoretically
predicted quasiparticle amplitudes of electrons and electron-hole
pairs in Eq.~\eqref{eq:Psi}.


Let us consider a generic quantum contact with spin-degenerate
transmission eigenvalues $T_n$ that are independent of energy. The
contact is driven by a voltage $V(t) = \Vdc + \Vac(t)$, where $\Vdc$
is a constant dc offset and $\Vac(t)$ is a periodic ac voltage
component with zero average and the period $T=2\pi/\omega$.
The cumulant generating function of the charge transfer statistics is
given by\cite{keeling_minimal_2006,vanevic_elementary_2008}
$\cS(\chi) = 2\sum_n \Tr \ln [1+ f_L(1-f_R)T_n(e^{i\chi}-1) \notag
+ (1-f_L) f_R T_n (e^{-i\chi}-1)]$.
Here, $f_{L(R)}$ are generalized electronic distribution functions in
the left (right) terminal which depend on two time or energy
arguments:
$f_L \equiv \tilde f = e^{-i\phi(t')} f_{\Vdc}(t'-t'')
e^{i\phi(t'')}$,
$f_R \equiv f(t'-t'')$.
Here, $\phi(t) = \int_0^t dt' e\Vac(t')/\hbar$ (hereafter $\hbar=1$),
$f(\cE)=1-\theta(\cE)$,  and $f_{\Vdc}(\cE)=f(\cE-e\Vdc)$, where
$\theta$ is the step function.
Function $\tilde f$ couples only energies which differ by an integer
multiple of $\omega$,
$\tilde f(\cE', \cE'') = \sum_k a_k a_{k+m-n}^* f(\cE' - k\omega -
e\Vdc)$,
where $\cE'=\epsilon+n\omega$, $\cE''=\epsilon+m\omega$
($0<\epsilon<\omega$), and $a_n = (1/T)\int_0^T dt e^{-i\phi(t)}
e^{in\omega t}$. In particular, diagonal components of $\tilde f$ are
given by
$\tilde f(\cE) = \sum_{n=-\infty}^\infty |a_n|^2
f(\cE-n\omega-e\Vdc)$.
Function $\tilde f(\cE',\cE'')$ is the generalized distribution
function of a driven quantum contact in energy representation, while
$\tilde f(\cE)$ is the stationary nonequilibrium electronic
distribution which is realized in the junction due to time-dependent
drive.

Next we obtain a decomposition of $\tilde f$ into single-electron and
electron-hole states. As shown in Ref.
\onlinecite{vanevic_elementary_2007},
the notion of single-electron and electron-hole pair excitations is
related to the eigenproblem of $\{h,\tilde h\}\equiv h\tilde h +
\tilde h h$, where $f = (1-h)/2$ and $\tilde f = (1- \tilde h)/2$.
For integer $e\Vdc/\omega = N$, there is an $N$-dimensional subspace
of $\{h,\tilde h\}$ spanned by $N$ {\em special} vectors that are
eigenvectors of {\em both} $h$ and $\tilde h$: $\tilde h \mb v = -
\mb v$, $h \mb v = \mb v$. This subspace corresponds to $N$ {\it
electrons} injected to the contact per voltage cycle. In addition,
the operator $\{h,\tilde h\}$ has a series of two-dimensional
subspaces that are spanned by vectors $\mb v_\ga$ and $\mb v_{-\ga}
\equiv h \mb v_\ga$ which are given by $h\tilde h \mb v_\ga =
e^{i\ga} \mb v_\ga$. The spaces $\spann(\mb v_{\ga_k}, \mb
v_{-{\ga_k}})$ correspond to the {\it electron-hole pairs} (labelled
by $k=1,2,\ldots$) created per voltage cycle with the probabilities
$p_k=\sin^2(\ga_k/2)$.

At zero temperature, $f$ and $\tilde f$ commute with $\{h,\tilde
h\}$, that is, they reduce in the single-electron and electron-hole
pair subspaces of $\{h,\tilde h\}$. For simplicity, we restrict our
consideration to the case $N = 1$ in which there is only one electron
injected per voltage cycle. The eigenproblem of $\{h, \tilde h\}$
defines a resolution of the identity $\ket{\mb v}\bra{\mb v} + \sum_k
\hat P_k = 1$, where $\ket{\mb v}\bra{\mb v}$ is the projector on the
single-electron state and $\hat P_k = \ket{\mb v_{\ga_k}}\bra{\mb
v_{\ga_k}} +\ket{\mb v_{-\ga{_k}}}\bra{\mb v_{-\ga_k}}$ are the
projectors on the electron-hole subspaces. This defines a
decomposition of $\tilde f$ into single-electron and electron-hole
contributions,
\begin{equation}\label{eq:ft-decomp-1}
\tilde f = \ket{\mb v}\bra{\mb v} + \sum_k \tilde f_k,
\end{equation}
where $\tilde f_k = \tilde f \hat P_k$. Similarly, for the completely
filled Fermi sea of the right lead we have $f=\sum_k f_k$ where $f_k
= f \hat P_k$. The single-electron state $\mb v$ is given by $\tilde
f \mb v = \mb v$ and $f \mb v = 0$.

The first term $\ket{\mb v}\bra{\mb v}$ in Eq.~\eqref{eq:ft-decomp-1}
describes a single-electron state injected to the contact while
$\tilde f_k$ describe the electron-hole pairs. By taking diagonal in
time components of $\ket{\mb v}\bra{\mb v}$ or $\tilde f_k$ we gain
information on the time dependence of single-electron and
electron-hole pair wave functions. Similarly, by taking diagonal in
energy components we gain information on the single-electron and
electron-hole pair contributions in the overall distribution
function.
Indeed, for Lorentzian pulses $V_{\rm Lor}(t)$ that carry a single
charge quantum per cycle, there are no additional electron-hole
excitations and the time-dependent probability of the single-electron
injection is proportional to the drive, $|\mb v(t)|^2 = eV_{\rm
Lor}(t)/\omega$. This is no longer true for a general time-dependent
drive where $|\mb v(t)|^2 \ne eV(t)/\omega$ due to the presence of
electron-hole pairs.

Before we proceed with the specific examples, let us bring $\tilde
f_k$ in a more transparent form. States $\mb v_\ga$ and $\mb
v_{-\ga}$ in general possess both positive and negative energy
components. We can make a rotation of the basis $\mb u_\pm = (\mb
v_\ga \pm \mb v_{-\ga})/\sqrt{2}$ in the subspace $\spann(\mb
v_{\ga}, \mb v_{-\ga})$ such that new basis vectors $\mb u_+$ and
$\mb u_-$ possess non-zero components only for $\cE > 0$ and $\cE <
0$, respectively. Using $h\mb v_\ga = \mb v_{-\ga}$ it is
straightforward to check that $f \mb u_+=0$ and $f \mb u_- = \mb
u_-$, as required. The projector $\hat P_k$ in the new basis reads
$\hat P_k = \ket{\mb u_+^k}\bra{\mb u_+^k} +\ket{\mb u_-^k}\bra{\mb
u_-^k}$. We obtain
\begin{equation}\label{eq:ftk-u}
\tilde f_k =
    p_k \ket{\mb u_+^k}\bra{\mb u_+^k}
    + q_k \ket{\mb u_-^k}\bra{\mb u_-^k}
    -i\sqrt{p_k q_k}\, \ket{\mb u_+^k}\bra{\mb u_-^k} + {\rm h.c.}
\end{equation}
and $f_k = \ket{\mb u_-^k}\bra{\mb u_-^k}$ ($q_k=1-p_k$).
Decomposition of $\tilde f$ in Eqs.~\eqref{eq:ft-decomp-1} and
\eqref{eq:ftk-u} enables us to identify the many-body electronic
state in the left lead created by the drive. We find that $\tilde
f(\cE',\cE'')= \bra{\Psi} \hat c^\dagger(\cE')\hat
c(\cE'')\ket{\Psi}$ where $\ket{\Psi}$ is given in
Eq.~\eqref{eq:Psi}. The right lead is assumed to be unperturbed and
serves as a reference.

The physical meaning of the amplitudes ${\rm u}_\pm$ is manifest in
Eq.~\eqref{eq:Psi} and can further be elaborated by taking diagonal
in energy components of $\tilde f_k$. For $\cE>0$ we find that
$\tilde f_k(\cE)=p_k |\mb u_+^k(\cE)|^2$. For $\cE<0$ it is more
convenient to consider the distribution of holes $\tilde f_k^{(h)}
\equiv \hat P_k - \tilde f_k$: $\tilde f_k^{(h)}(\cE)=p_k |\mb
u_-^k(\cE)|^2$. Therefore, $\mb u_+$ and $\mb u_-$ describe electrons
and holes generated in the system. This notion is also supported in
time domain with diagonal in time components $\tilde f_k(t) = |\mb
u_+^k(t)|^2$ and $\tilde f_k^{(h)}(t) = |\mb u_-^k(t)|^2$.
%
Apart from electron and hole states on the diagonal, $\tilde f_k$
contains also the off-diagonal terms proportional to
$\sqrt{p_k(1-p_k)}$ that are responsible for mixing of the two, see
Eq.~\eqref{eq:ftk-u}. Electron-hole pairs with $p_k\approx 0$ give no
contribution to the transport. On the other hand, for $p_k\approx 1$
the electron and the hole from a pair are practically decoupled from
each other (off-diagonal mixing terms vanish), cf.
Eq.~\eqref{eq:Psi}. This can also be seen in the cumulant generating
function which becomes a sum of independent electron and hole
contributions.\cite{vanevic_elementary_2008}

\begin{figure}[t]
\includegraphics[scale=1.23]{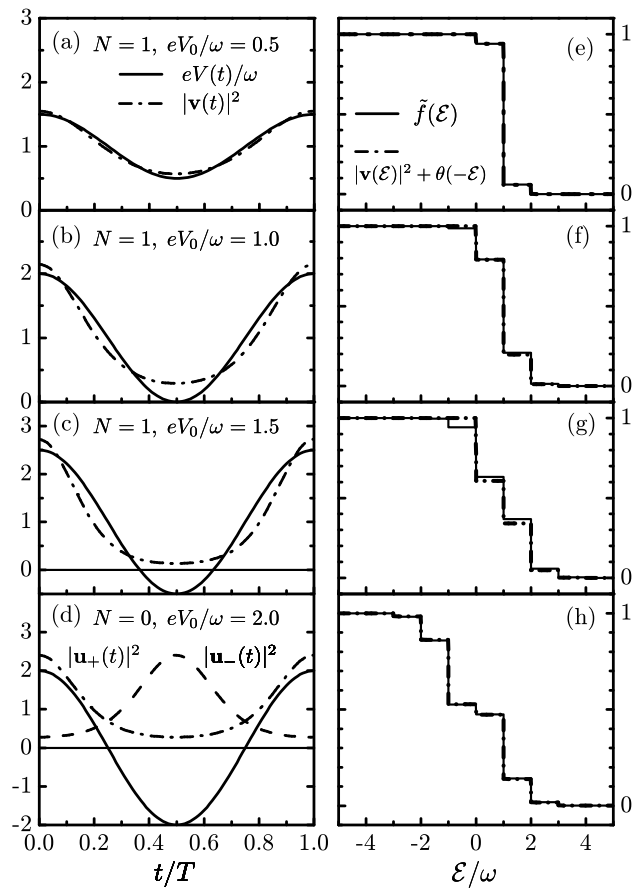}
\caption{\label{fig:wavepackets-fE-all} (a)--(d) Wave functions $|\mb
v(t)|^2$ and $|\mb u_\pm(t)|^2$ of electrons and electron-hole pairs
created by different voltage drives $V(t)$. The corresponding
nonequilibrium distribution functions are shown in the panels
(e)--(h).}
\end{figure}

Next we study single-electron and electron-hole pair states
for different voltage drives, see Fig.~\ref{fig:wavepackets-fE-all}.
Let us consider a harmonic drive $V(t) = \Vdc + V_0 \cos(\omega t)$
where the dc offset $N = 1$ is kept fixed while the amplitude $V_0$
of the ac component is varied. The drive is characterized by the
coefficients $a_n = J_n(eV_0/\omega)$ where $J_n$ are the Bessel
functions of the first kind. In addition to a single-electron state,
there are also electron-hole pairs created in the system and they
become more relevant for transport as the amplitude $V_0$ is
increased. For $eV_0/\omega \lesssim 2$ there is only one
electron-hole pair with $p_1 = (1/2) (\sum_{n=-\infty}^\infty |n+N|\;
|a_n|^2 - N)$ in addition to the single-electron state injected. For
$eV_0/\omega \ll 1$, the probability $p_1$ practically vanishes and
only a single-electron state remains. Time dependence of the
single-electron wave packet $|\mb v(t)|^2$ is shown in
Figs.~\ref{fig:wavepackets-fE-all}(a--c) in comparison to the voltage
drive $V(t)$. We find that for ac drive amplitudes much smaller than
dc voltage bias, the electron-hole pair creation is not effective and
the single-electron wave packet coincides with the drive, $|\mb
v(t)|^2\approx eV(t)/\omega$, see
Fig.~\ref{fig:wavepackets-fE-all}(a). For ac drive amplitudes
comparable to or larger than dc offset, the single-electron wave
packet differs from the drive, see
Figs.~\ref{fig:wavepackets-fE-all}(b,c). In that case the
electron-hole pairs become important and their wave packets together
with the single-electron one ensure $I(t)=G V(t)$, where
$G=(e^2/\pi)\sum_n T_n$. The wave functions $|\mb u_\pm(t)|^2$ of an
electron-hole pair are shown in Fig.~\ref{fig:wavepackets-fE-all}(d)
for the drive without dc bias ($N=0$) which does not create
single-electron states. The nonequilibrium distribution functions
$\tilde f(\cE)$ for the voltage drives at hand are shown in
Figs.~\ref{fig:wavepackets-fE-all}(e--h) (solid curves) together with
the approximations (dash-dotted curves) computed using the most
important single-electron or electron-hole pair states in
Figs.~\ref{fig:wavepackets-fE-all}(a--d). From
Figs.~\ref{fig:wavepackets-fE-all}(b,c,f,g) we find that
electron-hole pairs give a more significant contribution in time
domain than in $\tilde f(\cE)$. Indeed, while the single-electron
wave packets clearly differ from the voltage drive, the distribution
function is nevertheless to a good accuracy given by a
single-electron state, $\tilde f(\cE)\approx |\mb v(\cE)|^2 +
\theta(-\cE)$. This is because for small $p_k$, the electron-hole
pair functions $\tilde f_k$ in Eq.~\eqref{eq:ftk-u} have dominant
off-diagonal electron-hole mixing components (proportional to
$\sqrt{p_k}$) which do not contribute to diagonal in energy
distribution $\tilde f_k(\cE)$, while they do contribute to diagonal
in time distribution.

The time dependence of the electronic wave functions can be accessed
experimentally by studying current noise power in a setup where two
voltage drives with time shift $\tau$ are applied to the terminals,
$V_L(t) = \Vdc + V_0 \cos(\omega t)$ and $V_R(t) = V_L(t-\tau)$. This
can be viewed as the electronic analog of the optical Hong-Ou-Mandel
(HOM) experiment\cite{hong_measurement_1987} in which electron wave
packets
emitted from the terminals with time delay $\tau$ collide at the contact.%
\cite{dubois_integer_2013,grenier_single-electron_2011}
%
%
In the analog HOM experiment, $\Vdc$ is the static bias voltage
between the input and output ports, which is in our case only defined
with respect to virtual output terminals. Because of the gauge
invariance, our two-terminal setup is formally equivalent to the case
of the voltage $\delta V(t) = V_L(t)-V_R(t)$ applied only to the left
lead with the right lead unperturbed. The current noise power as a
function of the time delay $\tau$ then reads
$ S_2(\tau) = S_0 \sum_{n=-\infty}^\infty |n| J_n^2 [
(2eV_0/\omega)\sin(\omega\tau/2) ]$
where $S_0 = (e^2\omega/\pi) \sum_n T_n R_n$ with
$R_n=1-T_n$.\cite{note_Vdc}
To express $S_2(\tau)$ in terms of the overlap of the wave packets,
we proceed as follows. From the cumulant generating function $\cS$ we
find $ S_2(\tau) \propto \Tr [(f_L - f_R)^2] = \Tr[(f_L-f)^2 +
(f_R-f)^2 - 2(f_L-f)(f_R-f)]$
where we have introduced the distribution function $f$ of the
unperturbed Fermi sea. Here, the first two terms on the right-hand
side give the noise when the voltage is applied to one lead only
while the other lead is unperturbed. Both terms give the same
contribution to the noise $S_L = S_R = S_0 \sum_{n=-\infty}^\infty
|n+N| J_n^2(eV_0/\omega)$ independent of $\tau$. The term
$-\Tr[2(f_L-f)(f_R-f)]$ gives the noise suppression due to wave
packet overlap at the contact. The noise reads
\begin{equation}\label{eq:S2-overlap}
S_2(\tau)/S_0 = (S_L+S_R)/S_0 - 2C(\tau),
\end{equation}
where $C(\tau)$ is the overlap that we compute using
Eqs.~\eqref{eq:ft-decomp-1} and \eqref{eq:ftk-u}.
For $p_k\ll 1$, the dominant contribution in $C(\tau)$ is the overlap
of the single-electron wave functions $\mb v(t)$ and $\mb v'(t)\equiv
\mb v(t-\tau)$ injected from the leads,
\begin{equation}\label{eq:C-v}
C(\tau) \approx |\scp{\mb v}{\mb v'}|^2 = \left\vert \int_0^T
\frac{dt}{T}\; \mb v^*(t) \mb v(t-\tau) \right\vert^2.
\end{equation}
%
\begin{figure}[t]
\includegraphics[scale=1.345]{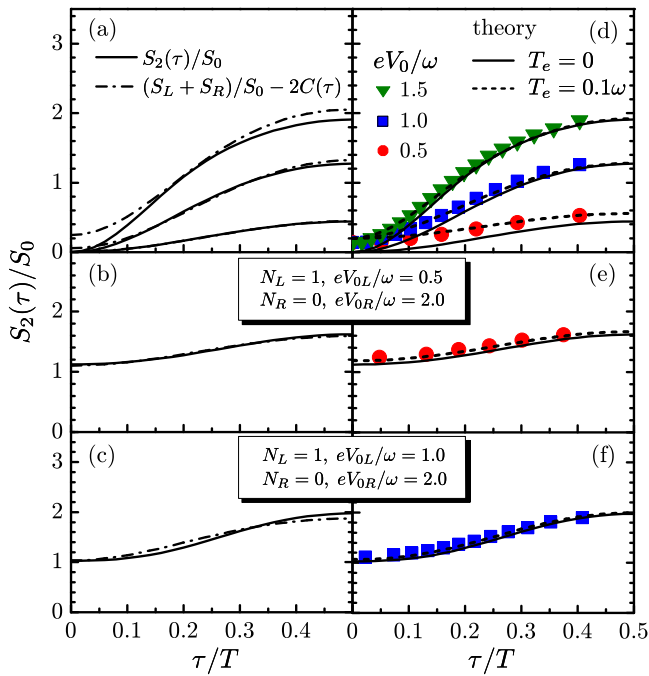}
\caption{\label{fig:overlaps-all-exp} Noise $S_2$ as a function of a
time delay $\tau$ between two harmonic signals applied at the leads:
(a) $N = 1$ and $eV_0/\omega = 1.5$, $1$, $0.5$ (solid lines, top to
bottom), (b) left lead: $N = 1$, $eV_0/\omega = 0.5$; right lead: $N
= 0$, $eV_0/\omega = 2$, (c) left lead: $N = 1$, $eV_0/\omega = 1$;
right lead: $N = 0$, $eV_0/\omega = 2$. Approximations for $S_2$
calculated using the overlaps $C(\tau)$ of the wave functions
depicted in Fig.~\ref{fig:wavepackets-fE-all} are shown for
comparison (dash-dotted lines). (d)--(f) $S_2(\tau)$ measured in a
tunnel junction at $\omega/2\pi = 20\,$GHz (symbols). Theoretical
results are shown for temperatures $T_e=0$ (solid lines) and
$T_e=0.1\omega$ (dashed lines).}
\end{figure}
%
The noise $S_2(\tau)$ is shown in Fig.~\ref{fig:overlaps-all-exp}(a)
together with the noise computed using the most important
single-electron wave packet overlap in Eq.~\eqref{eq:C-v}.
The corresponding voltage drives and the electron wave functions
are shown in Figs.~\ref{fig:wavepackets-fE-all}(a--c).

So far we have analyzed the single-electron wave packets. To probe
the electron-hole states we can use the voltages $V_L(t) = \Vdc +
V_{0L} \cos(\omega t)$ and $V_R(t) = V_{0R} \cos[\omega(t-\tau)]$
which create single-electron wave packets at the left lead and
electron-hole pairs at the right lead.
%
%
The current noise power in this case is given by $S_2(\tau) = S_0
\sum_{n=-\infty}^\infty |n+N|\; J_n^2[eV_0(\tau)/\omega]$, where
$V_0(\tau) = [V_{0L}^2+V_{0R}^2 -
2V_{0L}V_{0R}\cos(\omega\tau)]^{1/2}$. In terms of the overlap, the
noise is given by Eq.~\eqref{eq:S2-overlap} where $S_L = S_0 \sum_n
|n+N| J_n^2(eV_{0L}/\omega)$, $S_R = S_0 \sum_n |n|
J_n^2(eV_{0R}/\omega)$, and $C(\tau) \approx p_R |\scp{\mb v}{\mb
u'_+}|^2$ is the overlap between the single-electron state $\mb v$ at
the left lead and the electron part $\mb u'_+$ of the most dominant
electron-hole pair $\mb u_\pm'$ generated in the right lead
($p_R=0.630$). The noise $S_2(\tau)$ in this case is shown in
Figs.~\ref{fig:overlaps-all-exp}(b,c); the corresponding
single-electron and electron-hole wave functions are shown in
Figs.~\ref{fig:wavepackets-fE-all}(a,b,d).

To verify this picture, we have measured $S_2(\tau)$ in a tunnel
junction under harmonic excitation with $\omega/2\pi=20\,$GHz (see
Supplemental material).
Experimental results are shown in
Figs.~\ref{fig:overlaps-all-exp}(d--f), in agreement with the current
noise power obtained theoretically. This proves that single-electron
and electron-hole excitations in Eq.~\eqref{eq:Psi} can be created by
time-dependent voltage and accessed experimentally in a noise
correlation experiment which measures the overlap of the electronic
wave functions.

In conclusion, we have obtained the many-body electronic state
created by a time dependent drive of a quantum contact in terms of
single-electron and electron-hole quasiparticle excitations. We have
confirmed our theoretical predictions by probing the constituent
quasiparticle states in a HOM-type experiment on a tunnel junction.
The knowledge of the many-body state opens a way of engineering the
required time profile or energy distribution of single-electron and
electron-hole excitations. Since the electronic state in a conductor
determines the electromagnetic field it
generates,\cite{qassemi_quantum_2015} our work can be used to produce
non-classical states of electromagnetic field, such as squeezed or
entangled photonic states that have been observed recently.%
\cite{bednorz_nonsymmetrized_2013,*gasse_observation_2013a,forgues_experimental_2015}


\begin{acknowledgments}
We are grateful to Lafe Spietz for providing us with the sample. We
acknowledge Z. Radovi\'c and M. Aprili for valuable discussions. The
research was supported by the Serbian Ministry of Science Project
No.~171027, the bilateral project CNRS -- MESTD,
ANR-10-LABX-0039-PALM, ANR-11-JS04-006-01, DFG through SFB 767, the
German Excellence Initiative through CAP, and the Canada Excellence
Research Chairs program.
\end{acknowledgments}


\bibliography{ElectronStates}

\begin{thebibliography}{39}
\bibitem [{not()}]{note_Vdc}%
  \BibitemOpen
  \href@noop {} {}\bibinfo {note} {We note that the noise is independent of
  $\Vdc$ in our case. In a true HOM setup the finite dc-bias is applied between
  input and output leads. This will lead to the transmission of single-electron
  states and add a constant noise background independent of the time-dependent
  drive.}\BibitemShut {Stop}%
\end{thebibliography}

\clearpage

\begin{widetext}
\vspace*{1cm}
\begin{center}
\large\bf
Supplemental Material for\\
``Electron and electron-hole quasiparticle states in a driven quantum contact''
\end{center}
\vspace*{2cm}
\end{widetext}

\section{Experimental setup}

\begin{figure}
\begin{center}
\includegraphics[width=0.75\linewidth]{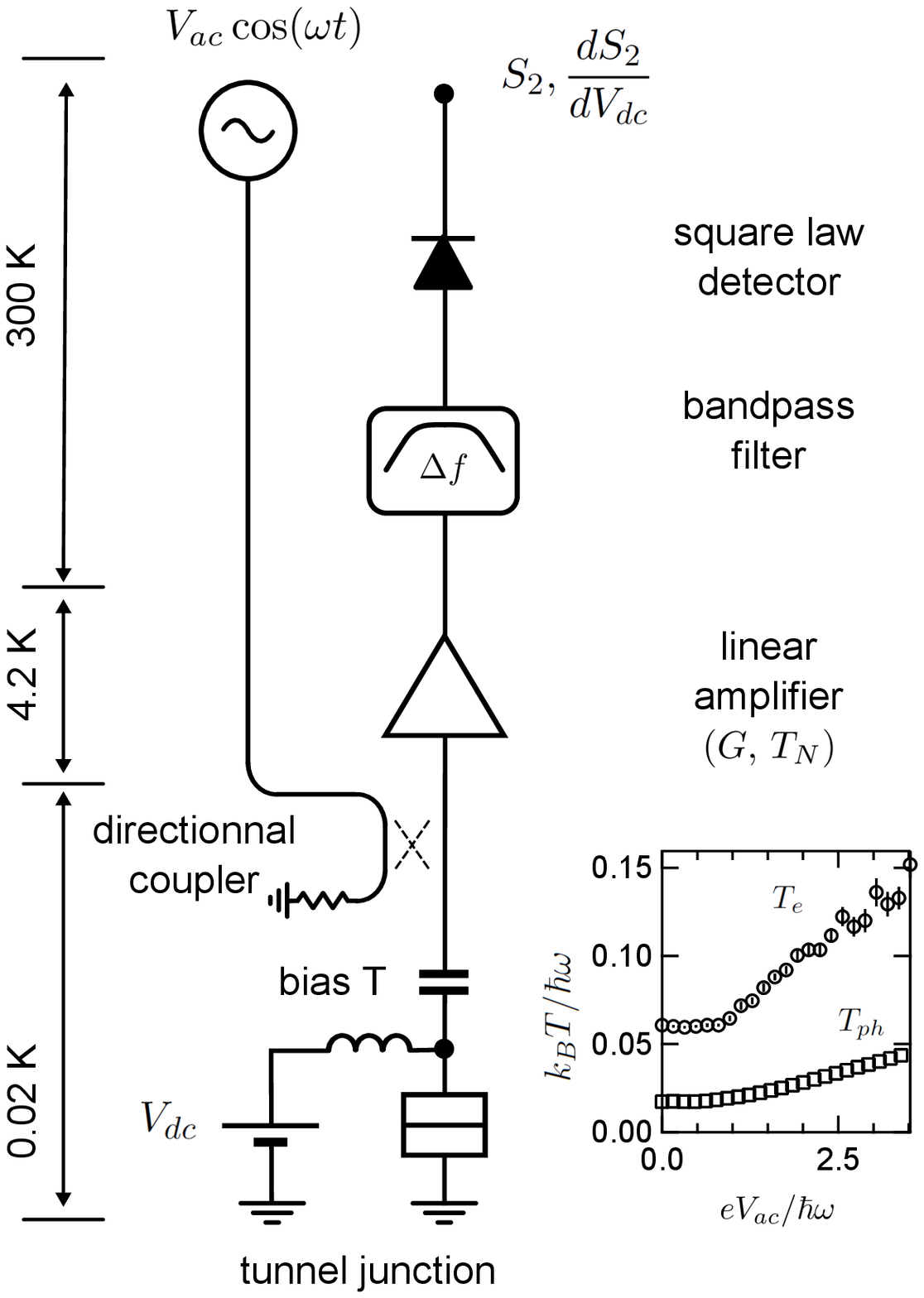}
\end{center}
\vspace{-0.5cm}
\caption{\label{fig:sup1}Experimental setup for the measurement of the current noise power in a tunnel junction under harmonic excitation.}
\end{figure}

In order to generate a single electron state by applying a periodic
voltage bias [$V(t)=V_{dc}+V_0 \cos (\omega t)$] on a quantum
contact, the experiment has to be performed in the quantum regime,
that is, $\hbar \omega \gg k_BT$ at low temperature $T \sim 100\,
\mathrm{mK}$ and high frequency $\omega/2\pi=20\, \mathrm{GHz}$. To
provide a good matching to the coaxial cable and avoid reflection of
the ac excitation, one use a $R_0 \simeq 50 \Omega$ Al/Al oxide/Al
tunnel junction as a quantum contact. Indeed, the time dependence of
single-electron and electron-hole wave functions in a driven quantum
contact does not depend either on the number nor on the transmission
of its conduction channels. The experiment is performed in a dilution
refrigerator where a $0.1 \,\mathrm{T}$ perpendicular magnetic field
is applied to turn the Al normal. A bias tee, sketched in Fig. \ref{fig:sup1} by
an inductor and a capacitor, allows to separate the dc bias voltage
applied on the junction from the high frequency part of the setup.
The ac excitation is imposed through a directional coupler and the
current fluctuations emerging from the sample are amplified by a low
noise cryogenic amplifier (noise temperature $T_N \sim 7 \, K$).
Current fluctuations are band-pass filtered between $\Delta f = 1.2 -
3 \, \mathrm{GHz}$ and the noise power density $S_2$ integrated over
the bandwidth of the filter is measured with a broadband square law
detectors as a function of dc bias voltage for different ac
excitations. The derivative of the noise $\partial
S_2/\partial(eV_{dc})$ is measured with an additional $77 \,
\mathrm{Hz}$, small voltage modulation and a usual lock-in detection.
The base temperature $T_{ph}$ of the dilution refrigerator
significantly increases with increasing ac voltage. It implies an
increase of the electronic temperature $T_e$ which is measured by
fitting the data of $\partial S_2/\partial(eV_{dc})$  (see inset in
Fig. \ref{fig:sup1}).

\section{Calibration procedure}

\begin{figure}[h!]
\begin{center}
\includegraphics[width=0.9\linewidth]{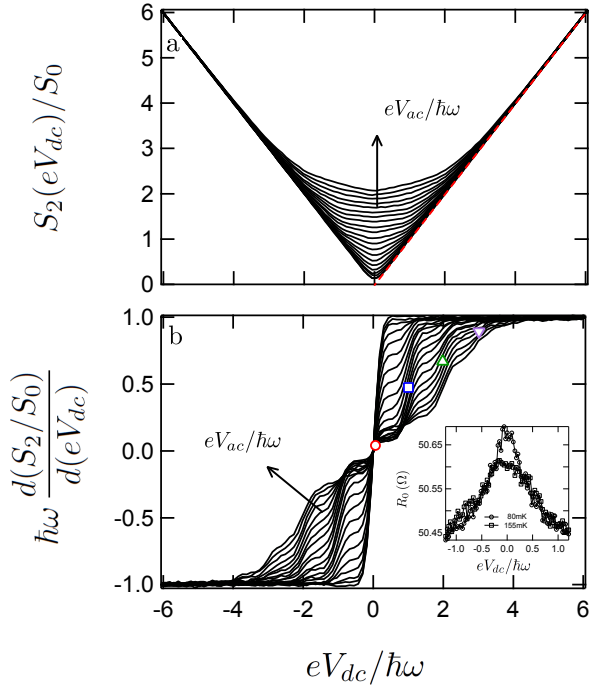}
\end{center}
\vspace{-0.5cm}
\caption{\label{fig:sup2}(a) Measured noise power density $S_2$ and (b) differential noise power density $dS_2/d(eV_{dc})$ for various levels of excitation $eV_{0}/\hbar \omega$. Arrows correspond to the sense of $eV_{0}/\hbar \omega$ increasing. {Inset:} Dynamic resistance measured at $80$ and $155 \, \rm{mK}$. Non-linearities disappear for $T_e>400 \, \rm{mK}$ and the resistance of the sample is $R_0 = 50.4 \,  \Omega$.}
\end{figure}

\begin{figure}[h!]
\begin{center}
\includegraphics[width=0.8\linewidth]{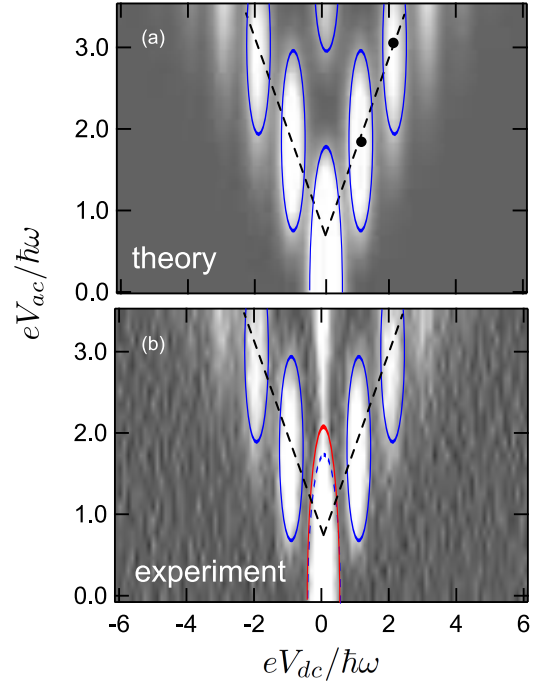}
\end{center}
\vspace{-0.5cm}
\caption{\label{fig:sup3}(a) Calculated and (b) measured second derivative of the photon-assisted noise $\partial^2(S_2/S_0)/\partial(eV_{dc})^2$ as a function of normalized dc bias $eV_{dc}/\hbar \omega$ and normalized ac bias $eV_{0}/\hbar \omega$. The black dots correspond to the maximum of $J_1^2(eV_{ac}/\hbar \omega)$ and $J_2^2(eV_{ac}/\hbar \omega)$.}
\end{figure}

\begin{figure}[h!]
\begin{center}
\includegraphics[width=0.8\linewidth]{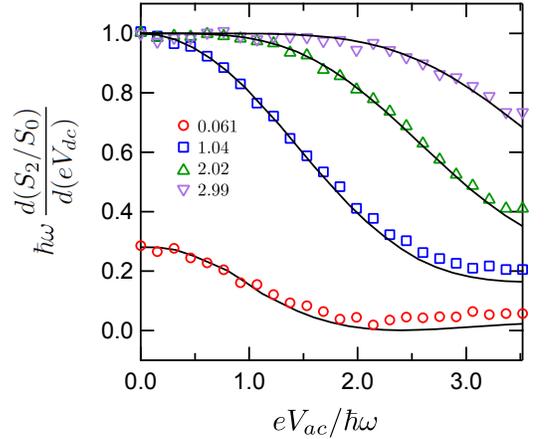}
\end{center}
\vspace{-0.5cm}
\caption{\label{fig:sup4}Calculated (markers) and fitted (solid lines) derivative of the photon-assisted noise $\partial(S_2/S_0)/\partial(eV_{dc})$ as a function of normalized ac bias $eV_{ac}/\hbar \omega$ at different dc bias: $eV_{dc}/\hbar \omega=0.061$, $1.04$, $2.02$ and $2.99$.}
\end{figure}

The current noise power associated to the collision of two
single-electron wave packets generated by two voltage drives
$V_L(t)=V_{dc}+V_0 \cos(\omega t)$ and $V_R(t)=V_L(t-\tau)$ is simply
given by the harmonic photon-assisted noise with $V_{ac}=2eV_0
\sin(\omega \tau/2)$:

$$S_2(eV_{dc},eV_{ac})=\sum_{n=-\infty}^{+\infty} J_n^2\left(\frac{eV_{ac}}{\hbar \omega} \right)\, \frac{R_0 (eV_{dc}+n\hbar \omega)}{\mathrm{th}\left(\frac{eV_{dc}+n\hbar \omega}{2k_BT_e}\right)},$$

\noindent
where $S_0= \hbar \omega /R_0$ and $T_e$ is the temperature of
electrons. Since we measure the amplified current noise power density
$G(S_2+2k_BT_N) \Delta f$ and we want to extract its absolute value
$S_2$, we need to calibrate the total gain $G$ of the measurement
setup (including the amplification and the cable attenuation), the
noise added by the amplifier $T_N$ and the coupling $eV_{ac}/\hbar
\omega$ between the sample and the excitation line. This calibration
has been done in two stages: (i) $G \, \Delta f$ and $T_N$ are
deduced from the high voltage dependence of the noise power which is
simply given by the classical shot noise: $S_2(eV_{dc}\gg eV_{ac},
k_BT_e)=eV_{dc}/R_0$ (see red dash line in Fig. \ref{fig:sup2}a). (ii) Figure \ref{fig:sup3}
shows the calculated (a) and measured (b) seconde derivative of the
photon-assisted noise $\partial^2(S_2/S_0)/\partial(eV_{dc})^2$. One
observes maxima corresponding to maxima of $J_n^2(eV_{ac}/\hbar
\omega)$ which are approximatively aligned with $eV_{ac}=(0.79 \pm
0.05)\times \hbar \omega + (1.06 \pm 0.01) \times eV_{dc}$ (black
dash line on Fig. \ref{fig:sup3}). These remarkable points are used to calibrate
ac coupling and our data are in very good agreement with the
theoretical predictions for $eV_{dc}/\hbar \omega \geq0.5$. However,
we notice a discrepancy for $eV_{dc}/\hbar \omega<0.5$ (red circle on
Fig. \ref{fig:sup3}b) which reveals a smaller coupling at low dc bias voltage. It
is attributed to non-linearities due to Coulomb blockade effects
appearing at low temperature and low bias voltage (see inset on Fig.
\ref{fig:sup2}b). Indeed, in spite of a small effect on the resistance ($\delta R
/R_0 \sim 1 \%$) the impedance mismatch can be emphasized by
interference effects. To properly calibrate the ac coupling, we have
plotted on Fig. \ref{fig:sup4} the derivative of the photon-assisted noise
$\frac{\partial(S_2/S_0)}{\partial(eV_{dc})}$ as a function of
normalized ac bias $eV_{ac}/\hbar \omega$. The fit gives an ac
coupling $6 \, \%$ lower at low bias voltage ($eV_{dc}/\hbar
\omega<1$) than high bias voltage ($eV_{dc}/\hbar \omega \geq 1$).

\mbox{}

\newpage
\mbox{}

\mbox{}

\end{document}